\documentclass[12pt]{article}
\usepackage{amsmath}
\usepackage{amssymb}
\usepackage{slashed}
\usepackage{graphicx}
\usepackage{subfigure}
\usepackage{cite}
\usepackage{url}
\usepackage[small]{caption}
\begin{document}
\baselineskip 0.5cm

\def\bra#1{\langle #1 |}
\def\ket#1{| #1 \rangle}
\def\inner#1#2{\langle #1 | #2 \rangle}

\begin{titlepage}

\begin{flushright}
\end{flushright}

\vskip 1.8cm

\begin{center}
{\Large \bf Why Firewalls Need Not Exist}

\vskip 0.7cm

{\large Yasunori Nomura and Nico Salzetta}

\vskip 0.4cm

{\it Berkeley Center for Theoretical Physics, Department of Physics,\\
 University of California, Berkeley, CA 94720, USA}

\vskip 0.1cm

{\it Theoretical Physics Group, Lawrence Berkeley National Laboratory, 
 Berkeley, CA 94720, USA}

\vskip 0.8cm

\abstract{The firewall paradox for black holes is often viewed as 
 indicating a conflict between unitarity and the equivalence principle. 
 We elucidate how the paradox manifests as a limitation of semiclassical 
 theory, rather than presents a conflict between fundamental principles. 
 Two principal features of the fundamental and semiclassical theories 
 address two versions of the paradox:\ the entanglement and typicality 
 arguments.  First, the physical Hilbert space describing excitations 
 on a fixed black hole background in the semiclassical theory is 
 exponentially smaller than the number of physical states in the 
 fundamental theory of quantum gravity.  Second, in addition to the 
 Hilbert space for physical excitations, the semiclassical theory 
 possesses an unphysically large Fock space built by creation and 
 annihilation operators on the fixed black hole background.  Understanding 
 these features not only eliminates the necessity of firewalls but also 
 leads to a new picture of Hawking emission contrasting pair creation 
 at the horizon.}

\end{center}
\end{titlepage}

\section{Introduction}
\label{sec:intro}

Ever since the discovery of the thermodynamic behavior 
of black holes~\cite{Bekenstein:1973ur,Bardeen:1973gs,Hawking:1974rv}, 
we have been searching for a deeper structure of spacetime and gravity 
beyond that described by general relativity.  Its exploration, however, 
has repeatedly led to confusions involving fundamental principles 
such as unitarity of black hole evolution and smoothness of their 
horizons~\cite{Hawking:1976ra,'tHooft:1990fr,Susskind:1993if,%
Almheiri:2012rt}; see, e.g., Refs.~\cite{Preskill:1992tc,Harlow:2014yka} 
for reviews.  In this regard, the latest major puzzle is the firewall 
paradox~\cite{Almheiri:2012rt,Almheiri:2013hfa,Marolf:2013dba}, which 
asserts that unitarity of black hole evolution as viewed from the 
exterior is inconsistent with smoothness of the horizon, assuming 
that the semiclassical theory is valid away from the stretched 
horizon.  It has been argued that the most likely implication of 
this is that an infalling observer encounters drama at the horizon, 
so that there is no such thing as the interior spacetime, at least 
for an old black hole in which the information retrieval process 
is operative~\cite{Page:1993wv}.

In this paper, we elucidate how the firewall paradox may manifest 
as a limitation of the semiclassical theory, rather than presents 
a conflict between fundamental principles.  We do this by illustrating 
how an interpretation of the semiclassical theory undermines some of the 
assumptions that went into the arguments of Refs.~\cite{Almheiri:2012rt,%
Almheiri:2013hfa,Marolf:2013dba}.  In fact, by using this understanding 
of the paradox we can explore the Hilbert space structure of matter 
and spacetime in the fundamental theory of quantum gravity.  While 
the picture we present is already implicit in more complete treatments 
of evaporating black holes in Refs.~\cite{Nomura:2014woa,Nomura:2014voa,%
Nomura:2015fja}, we find it useful to explicitly extract the features 
responsible for avoiding the existence of firewalls.  In particular, 
the following aspects of the fundamental and semiclassical theories 
play key roles:
\begin{itemize}
\item
The number of physical configurations representing semiclassical 
excitations, i.e.\ the configurations that are physically realized and 
which the operators in the semiclassical theory can discriminate, is 
much (exponentially) smaller than the number of physical states in 
the fundamental theory of quantum gravity.  This implies that in the 
fundamental theory, or the ``dual field theory,'' the same semiclassical 
operators can be realized in exponentially many different ways.  In 
other words, the actions of these operators are defined only on a subset 
of the whole degrees of freedom in the fundamental theory.
\item
The semiclassical theory possesses a (formally infinitely) large Hilbert 
space constructed as the Fock space associated with the creation and 
annihilation operators on a fixed black hole background.  This is because 
the effect of the excitations on the spacetime background is ignored in 
the semiclassical theory.  The finite number of independent configurations 
for the physical semiclassical excitations are mapped into this Hilbert 
space.  In other words, the elements of the Hilbert space outside the 
image of this map are unphysical, and as such, they do not exist in 
the corresponding dual field theory.
\end{itemize}
We argue that these two features are responsible for addressing 
the two representative arguments for firewalls:\ the entropy and 
typicality arguments.  After reviewing the firewall paradox in 
Section~\ref{sec:firewall} and presenting our view on the semiclassical 
approximation in Section~\ref{sec:semiclassical}, we refute the 
typicality and entanglement arguments in Sections~\ref{sec:refute-typ} 
and \ref{sec:refute-ent}, respectively.  In Section~\ref{sec:emission}, 
we present the picture of Hawking emission~\cite{Nomura:2014woa,%
Nomura:2014voa} implied by these analyses.

For simplicity, we present our analysis for a Schwarzschild black hole 
in 4-dimensional spacetime, although we do not expect difficulty in 
extending to other cases.  Throughout the paper, we do not discriminate 
the Planck length, $l_{\rm P}$, and the string length, but they can 
be straightforwardly separated if needed.  We use natural units in 
which $\hbar = c = l_{\rm P} = 1$, unless otherwise stated.

\section{The Firewall Paradox}
\label{sec:firewall}

Recall that the firewall arguments asserted that the complementarity 
picture~\cite{Susskind:1993if} was not enough to answer the black 
hole information problem.  What is the complementarity picture? 
Despite what Hawking considered long ago~\cite{Hawking:1976ra}, we 
now do not think that the black hole formation and evaporation process 
violates unitarity, at least from the viewpoint of a distant observer 
(based mainly on gauge/gravity duality~\cite{Maldacena:1997re}). 
This, however, raises the black hole ``information cloning 
paradox''~\cite{Preskill:1992tc}:\ the complete information 
about an object fallen into a black hole seems to reside {\it both} 
in late Hawking radiation {\it and} in the interior region, violating 
the no-cloning theorem in quantum mechanics.  The complementarity 
picture was proposed to address this paradox.  The basic idea was 
that no one can be distant and infalling observers {\it at the same 
time}, physically obtaining the information both from Hawking radiation 
and the fallen object.  The hope was that when one restricts the 
application of the classical spacetime picture to a causal patch 
(i.e.\ the spacetime region which a single observer, represented 
by a timelike geodesic, can access), semiclassical field theory 
still gives a good local description of physics.

A key point of the firewall arguments was that a paradox similar to 
the information cloning one could be formulated within a single causal 
patch.  The argument presented originally in Ref.~\cite{Almheiri:2012rt} 
goes as follows.  Consider an outgoing mode $B$ localized in the black 
hole zone region, $r < r_{\rm z} \simeq 3M$, which corresponds to Hawking 
radiation just emitted from the stretched horizon at $r = r_{\rm s} 
= 2M + O(1/M)$.  Here, $r$ is the Schwarzschild radial coordinate. 
For a sufficiently old black hole, unitarity requires this mode 
to be entangled with a mode representing Hawking radiation emitted 
earlier~\cite{Page:1993wv}.  On the other hand, according to semiclassical 
field theory, the smoothness of the horizon requires that any mode in 
the zone region, including $B$, must be entangled (almost maximally) 
with the pair mode inside the horizon~\cite{Unruh:1976db}.  These two 
statements cannot be reconciled.  A single mode $B$ cannot be entangled 
with two different modes, i.e.\ the earlier Hawking radiation mode 
(at $r > r_{\rm z}$) and the interior mode (at $r < r_{\rm s}$), 
since it would violate strong subadditivity of the entropy, entailing 
the information cloning.  We call this argument for firewalls the 
{\it entropy argument}.

Another argument was subsequently put forward using a putative 
map between a mode in semiclassical field theory (e.g.\ $B$ above) 
and an operator in the dual field theory.  The most sophisticated 
version~\cite{Marolf:2013dba} calculates the average of the number 
operator, $\hat{a}^\dagger \hat{a}$, in the dual field theory over 
states having energies in a chosen range, with $\hat{a}$ corresponding 
to an infalling mode $a$ in the bulk.  It was claimed that the resulting 
number is at least of order unity, $\bar{N}'_a \gtrsim O(1)$, because 
one can choose a basis for these states such that they are all 
eigenstates of the number operator $\hat{b}^\dagger 
\hat{b}$ with $\hat{b}$ corresponding to an exterior mode localized 
in the zone region (and because the expectation value of $\hat{a}^\dagger 
\hat{a}$ in any eigenstate of $\hat{b}^\dagger \hat{b}$ is at least 
of order unity).  This would imply that the expectation value of 
$\hat{a}^\dagger \hat{a}$ is of order unity or larger in a typical 
black hole state, i.e.\ most black hole states have firewalls. 
We call this argument the {\it typicality argument}.

The firewall paradox refers to a set of arguments indicating a conflict 
between unitarity of black hole evolution and smoothness of the horizon 
implied by the equivalence principle, formulated within a single causal 
patch.  The two arguments described above represent the most well 
developed among those formulated so far.

\section{Semiclassical Approximation}
\label{sec:semiclassical}

What is the semiclassical approximation?  Answering this question 
accurately is a key to resolving the firewall paradox.  Here we present 
a picture focusing on the relation between the Hilbert spaces of 
fundamental quantum gravity and semiclassical theory.  This picture 
builds on earlier work of one of the authors (Y.N.) with Sanches 
and Weinberg~\cite{Nomura:2013lia,Nomura:2014yka,Nomura:2014woa,%
Nomura:2014voa,Nomura:2015fja}.

\begin{figure}[t]
\centering
  \includegraphics[clip,width=.9\textwidth]{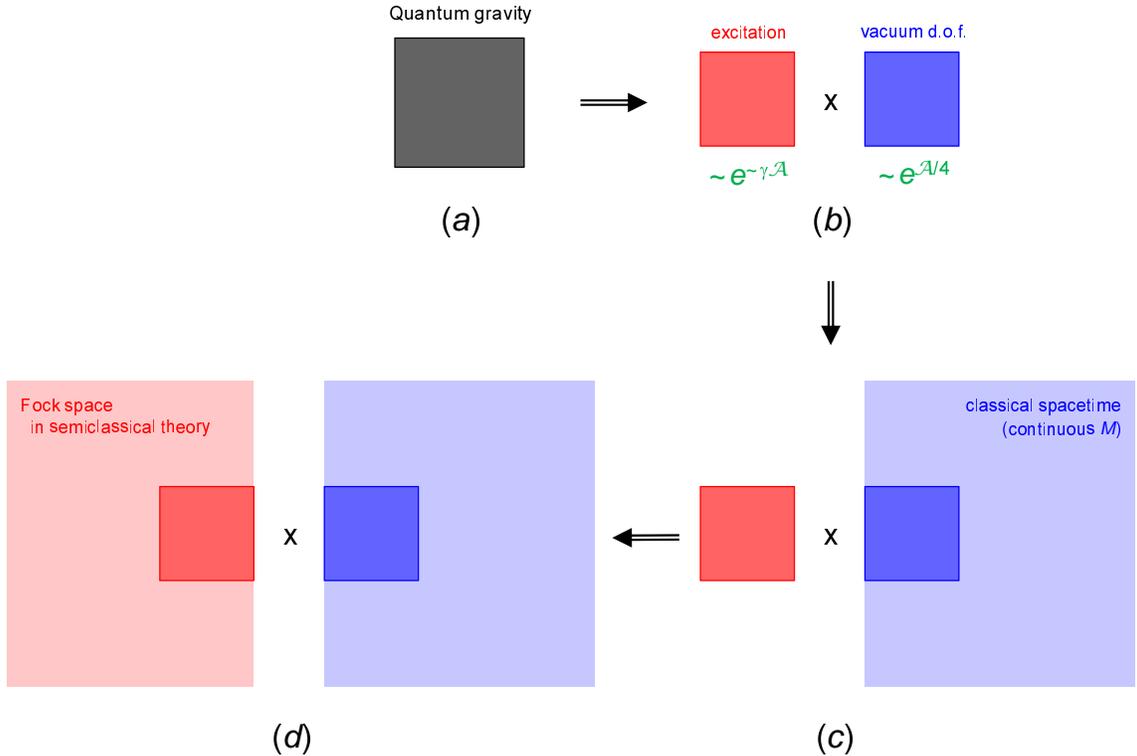}
\caption{Construction of the semiclassical approximation requires 
 splitting the physical degrees of freedom in quantum gravity, (a), 
 into the degrees of freedom associated with the black hole vacuum 
 (vacuum degrees of freedom) and excitations around it, (b).  The 
 vacuum degrees of freedom are then classicalized, (c), which creates 
 a large (fictitious) Hilbert space:\ the Fock space of creation and 
 annihilation operators on the resulting classical background, (d).}
\label{fig:semicl-1}
\end{figure}

Consider a set of quantum states representing a dynamical black hole 
of mass $M$ and its zone region, $r < r_{\rm z}$.  Here, we have 
adopted the Schr\"{o}dinger picture; in the Heisenberg picture 
this corresponds to considering a set of quantum states which have 
a black hole of mass $M$ at a fixed location at a fixed time, with 
the region outside the zone being unexcited.  The first step toward 
constructing the semiclassical approximation is to split the degrees 
of freedom represented by this set into those associated with the 
black hole ``itself'' and excitations around it.  According to the 
standard entropy argument, the number of independent black hole states 
without an excitation is
\begin{equation}
  {\cal N}_{\rm vac} \sim e^{\frac{1}{4}{\cal A} + O({\cal A}^p; p < 1)},
\label{eq:N_vac}
\end{equation}
where ${\cal A} = 16\pi M^2 \gg 1$ is the area of the black hole, and 
from now on we suppress possible higher order corrections in $1/{\cal A}$ 
in the exponents in analogous expressions.  The number of possible 
configurations for the excitations is expected to be
\begin{equation}
  {\cal N}_{\rm exc} \sim e^{\gamma {\cal A}};
\label{eq:N_exc}
\end{equation}
see, e.g., Ref.~\cite{'tHooft:1984re}.  Here, the coefficient $\gamma$ 
satisfies the holographic bound~\cite{'tHooft:1993gx}, $\gamma < 
(r_{\rm z}/4M)^2-1/4$.%
\footnote{In Refs.~\cite{Nomura:2014woa,Nomura:2014voa,Nomura:2015fja}, 
 it was stated that the number of physical configurations for the 
 excitations around the black hole is $\ln {\cal N}_{\rm exc} \sim 
 {\cal A}^q$ with $q < 1$, ignoring the effect of the redshift of the 
 Schwarzschild geometry.  Including this effect, the number of possible 
 configurations is rather $\ln {\cal N}_{\rm exc} \sim {\cal A}$. 
 This does not affect the basic conclusion.  Important points are 
 that $\ln {\cal N}_{\rm exc}$ is finite and that there are large 
 number of degrees of freedom, $\ln {\cal N}_{\rm vac}$, that cannot 
 be probed by operators in the semiclassical theory.}
Since the total number of quantum states is
\begin{equation}
  {\cal N} \approx {\cal N}_{\rm exc} {\cal N}_{\rm vac},
\label{eq:N}
\end{equation}
the physical Hilbert space describing excitations around a fixed black 
hole background is exponentially smaller than that of the whole quantum 
gravitational degrees of freedom, ${\cal N}_{\rm exc} \ll {\cal N}$.%
\footnote{A similar conclusion has also been reached in 
 Ref.~\cite{Almheiri:2014lwa} in the context of the AdS/CFT 
 correspondence.}
This first step is depicted as $(a) \rightarrow (b)$ in 
Fig.~\ref{fig:semicl-1}.

The next step is to ``classicalize'' the degrees of freedom corresponding 
to ${\cal N}_{\rm vac}$, which were called the vacuum degrees of freedom 
in Refs.~\cite{Nomura:2014yka,Nomura:2014woa,Nomura:2014voa} because they 
are associated with the black hole vacuum state in semiclassical theory. 
This step consists of two processes.  First, we must formally make 
the number of black hole degrees of freedom infinite as depicted as 
$(b) \rightarrow (c)$ in Fig.~\ref{fig:semicl-1} (although we will 
see later how semiclassical theory ``corrects'' this to represent 
phenomena associated with finite ${\cal N}_{\rm vac}$).  This can be 
understood by analyzing the origin of the Bekenstein-Hawking entropy, 
$\ln {\cal N}_{\rm vac} = {\cal A}/4$.  The quantum uncertainty principle 
implies that a dynamical black hole of mass $M$ has an energy uncertainty 
of $\varDelta E \approx \varDelta M \approx O(1/M)$ and, with the 
position uncertainty of order the quantum stretching of the horizon 
$\varDelta r \approx O(1/M)$, a momentum uncertainty of $\varDelta p 
\approx O(1/M)$.%
\footnote{The energy and momentum here refer to those as measured in 
 the asymptotic region.  The energy uncertainty, therefore, is given 
 by $\varDelta E \approx 1/\varDelta t$, where $\varDelta t$ is the 
 characteristic timescale for the change of the black hole state in 
 Schwarzschild time.  Assuming that the relevant timescale is the 
 Planck time as measured locally at the stretched horizon, we obtain 
 $\varDelta t \approx O(M)$.  This is indeed the timescale for 
 Hawking emission.}
The finiteness of the Bekenstein-Hawking entropy means that there are 
only a finite number of independent quantum states, $e^{{\cal A}/4}$, 
within these uncertainties.  On the other hand, classically, the number 
of independent states in this range is infinite, labeled by a {\it 
continuous} number $M$ (even ignoring the momentum uncertainty).%
\footnote{This is a standard phenomenon in the relation between 
 quantum and classical mechanics.  For example, the number of independent 
 states of a harmonic oscillator in a fixed energy interval is finite 
 in quantum mechanics (labeled by a discrete number for the levels) 
 while it is infinite in classical mechanics (labeled by a continuous 
 amplitude).}
Regarding the background spacetime as classical, therefore, amounts to 
enlarging the number of possible vacuum states to infinity.  This can 
also be seen from the fact that $\ln {\cal N}_{\rm vac}$ is written as 
${\cal A} c^3/4 l_{\rm P}^2 \hbar$ when $\hbar$, $c$, and $l_{\rm P}$ 
are restored, which becomes infinite for $\hbar \rightarrow 0$.%
\footnote{This implies that it is inaccurate to say that a classical 
 black hole loses exponentially many quantum hairs of a quantum black 
 hole.  The corresponding ``hairs'' for a classical black hole is 
 the mass parameter $M$.}

The second process is to ignore the backreaction, i.e.\ the effect 
of excitations on the (now classical) spacetime.  This comes with 
a major ``side effect'':\ since the effect of excitation degrees 
of freedom on the vacuum degrees of freedom is ignored, the resulting 
theory---semiclassical theory---allows for having a much larger (formally 
infinite) number of excitations on a fixed spacetime background.  In 
semiclassical field theory, this manifests itself as the fact that the 
Fock space built by creation and annihilation operators on the background 
is much larger than the actual Hilbert space for physical excitations; 
see (d) in Fig.~\ref{fig:semicl-1}.  In other words, the physical Hilbert 
space for the excitations is much smaller than what is naively implied 
by the Fock space in semiclassical field theory; by design, the semiclassical 
approximation is valid only for a very ``small'' number of excitations, 
of order $\ln {\cal N}_{\rm exc}$ or smaller.

At this point of the construction, the resulting theory seems fairly 
``superficial.''  The effect of excitations (matter and radiation) on 
spacetime is not automatically included---the only way to incorporate 
it is to solve the classical Einstein equation with a given configuration 
of the excitations (often taken as the quantum expectation value of 
the energy-momentum tensor) and adopt the resulting spacetime as the 
background.  The entropy of the black hole is formally infinite, so 
its temperature is zero---the black hole background exists eternally. 
However, the semiclassical approximation is actually more clever. 
It inherits some features reflecting the basic structure of the true 
physical degrees of freedom and their interactions, which allowed 
Hawking to discover the renowned black hole emission effect.

\begin{figure}[t]
\centering
  \includegraphics[clip,width=1\textwidth]{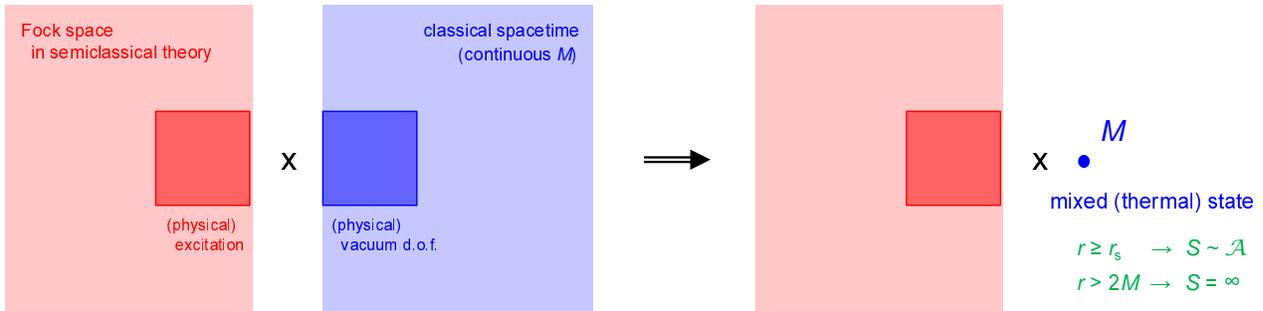}
\caption{Semiclassical theory as viewed from an external observer 
 encodes possible black hole microstates as the von~Neumann entropy 
 associated with the mixed black hole vacuum state.  By integrating 
 the entropy density associated with the local temperature from the 
 stretched horizon (Schwarzschild horizon) to the edge of the zone, 
 we obtain an entropy that scales as the area of the black hole (infinity), 
 which corresponds to the number of black hole microstates in the 
 quantum (classical) theory.}
\label{fig:semicl-2}
\end{figure}

Suppose we describe the system from the viewpoint of an external observer. 
If we want to describe the entire history of black hole evolution, we 
need to consider the whole time-dependent background from formation 
to evaporation, but if we are interested only in the black hole emission 
process, then we may consider a black hole background of mass $M$, which 
may be viewed as eternal at the semiclassical level~\cite{Unruh:1976db}. 
As we have discussed, the fact that the static approximation for the 
black hole is valid only for $\varDelta t \lesssim M$ implies that the 
state must have an uncertainty $\varDelta E \gtrsim 1/M$, so when we 
say a black hole of mass $M$ we are actually considering an ensemble 
of black holes of masses in the range $M \pm O(1/M)$.  Semiclassical 
field theory encodes this fact such that the black hole vacuum state 
is a mixed (thermal) state.  While this state is unique for a given 
$M$, the von~Neumann entropy of the state is nonzero, reflecting the 
fact that the black hole microstate in the fundamental theory is not 
unique (so with this procedure, black holes of slightly different masses 
in the range $M \pm O(1/M)$ need no longer be regarded as different). 
This is depicted in Fig.~\ref{fig:semicl-2}. 
Note that by integrating the thermal entropy density calculated using 
the local temperature
\begin{equation}
  T(r) = \frac{1}{8\pi M} \frac{1}{\sqrt{1-2M/r}},
\label{eq:T_r}
\end{equation}
from the stretched horizon, $r = r_{\rm s}$, to the edge of the zone, 
$r = r_{\rm z}$, we indeed obtain an entropy that scales as the area 
of the black hole, $S \sim {\cal A}$.  If we do not take into account 
the quantum stretching and integrate the entropy density from the 
Schwarzschild horizon, $r = 2M$, to the edge of the zone, $r = r_{\rm z}$, 
then we obtain $S = \infty$ consistently with the fact that the black 
hole entropy is infinite in the classical theory.

As we will discuss in more detail later, the thermal nature of the black 
hole vacuum state not only reflects the number of independent black hole 
microstates in the fundamental theory, but also encodes interactions of 
the black hole vacuum degrees of freedom with the rest of the degrees 
of freedom, e.g.\ field theory modes outside the zone, $r \gtrsim 
r_{\rm z}$.  Another important point here is that the number of physical 
excitations one can have on the black hole background is finite, of order 
$\ln {\cal N}_{\rm exc} \sim {\cal A}$.  The Hilbert space representing 
these excitations must be embedded into the infinitely large Fock space 
that one can formally build on the fixed black hole background.

\section{Refutation---The Typicality Argument}
\label{sec:refute-typ}

The picture in the previous section tells us how the typicality argument 
for firewalls can be flawed.  An important point is that the ``map'' 
from the physical Hilbert space of the fundamental theory to the Fock 
space of the semiclassical theory with a fixed black hole background 
is not one-to-one.  In particular, all the unexcited black hole 
microstates look exactly the same as probed by the operators in 
the semiclassical theory.  This occurs because these operators 
do not probe the vacuum degrees of freedom, i.e.\ the degrees of 
freedom in the right half in (c) and (d) of Fig.~\ref{fig:semicl-1} 
and the left panel of Fig.~\ref{fig:semicl-2}.  This implies that in 
the dual field theory, there are exponentially many different ways to 
represent the bulk semiclassical operators, which differ in actions 
on the degrees of freedom other than the excitation degrees of 
freedom.  Said differently, the actions of these operators are 
defined only on a subset (excitation) of the whole degrees of freedom 
(excitation~$+$~vacuum).  In Refs.~\cite{Nomura:2013lia,Nomura:2014yka,%
Nomura:2014woa,Nomura:2014voa}, this fact was referred to as that 
the semiclassical picture is obtained after coarse-graining the 
degrees of freedom associated with the Bekenstein-Hawking entropy.

Consider the creation and annihilation operators, $b^\dagger$ and $b$, 
corresponding to a mode localized outside the stretched horizon.  In 
terms of these operators, all the black hole vacuum states appear as 
a unique, thermal state.  The situation is analogous in the dual field 
theory.  In terms of the dual field theory operators $\hat{b}^\dagger$ 
and $\hat{b}$, which are the images of $b^\dagger$ and $b$, all the black 
hole vacuum states appear as the unique thermal state.  It is then clear 
that the average over all the $\hat{b}^\dagger \hat{b}$ eigenstates 
considered in Ref.~\cite{Marolf:2013dba} is irrelevant to the discussion 
on the smoothness of the horizon for the black hole vacuum states---all 
the states in the average are the same thermal state in terms of 
$\hat{b}^\dagger$ and $\hat{b}$ even with a finite width of energy 
range of order $\varDelta M$ in which the average is taken.  We stress 
that the thermal state in question should not be viewed as a statistical 
ensemble of states that look different as probed by the $\hat{b}^\dagger$ 
and $\hat{b}$ operators, as would be the case if the system were in 
thermal equilibrium in the usual sense.  This state is intrinsically 
mixed from the perspective of the semiclassical operators, and has 
the correct entanglement structure when the state is purified using 
the ``mirror'' modes $\tilde{b}$~\cite{Unruh:1976db,Israel:1976ur} 
at the semiclassical level.

What about the excited states?  One might think that if we take the 
average of Ref.~\cite{Marolf:2013dba} over {\it all} the black hole 
(not necessarily vacuum) states, one can take the basis of the states 
to be eigenstates of the number operator for one of the exterior modes, 
$b$, leading to the conclusion that a typical state must have firewalls. 
This is, however, not necessarily the case because the number of 
physical states is finite, $\sim {\cal N}$, so that the map from 
the physical states to the Fock space of the semiclassical theory 
may not be onto.  In particular, one may assume that the physical 
states correspond to states in which {\it infalling} modes $a$ (and/or 
excitations on the ``horizon'' as viewed from an infalling reference 
frame~\cite{Nomura:2014voa}) are excited.  In the distant description, 
this corresponds to states in which, when the vacuum state is purified 
using the mirror modes $\tilde{b}$, excitations preserve entanglement 
between $b$ and $\tilde{b}$ necessary to ensure the smoothness of 
the horizon.  For example, a state in which an $a$ mode is excited 
is described effectively as a state in which the thermal state is 
modulated by a linear combination of the $b$, $b^\dagger$, $\tilde{b}$, 
and $\tilde{b}^\dagger$ operators as implied by the Bogoliubov 
transformation between the $a$ and $b, \tilde{b}$ operators (although 
the full description may require intrinsically stringy effects 
because of a large boost between the infalling and distant reference 
frames~\cite{Nomura:2014voa,Susskind:1993aa}).  While we have not 
proved the assumption made here, we do not find a reason why it 
is impossible.

The picture described above implies that the average considered in 
Ref.~\cite{Marolf:2013dba} over the eigenstates of $b^\dagger b$---or 
its putative map $\hat{b}^\dagger \hat{b}$---corresponds to taking the 
average over an unphysically large Hilbert space, depicted as the light 
shaded (pink) region in the left half in (d) of Fig.~\ref{fig:semicl-1} 
and the two panels of Fig.~\ref{fig:semicl-2}.  One might think that 
if we impose a simple ultraviolet cutoff, e.g., the upper bound on 
the local energy of a $b$ quantum, then (the logarithm of) the number 
of states involved in the average becomes the right order, keeping 
the firewall argument.  This is, however, not the correct way to 
implement the cutoff.  The set of the physical states, i.e.\ the 
states that actually exist in the dual field theory, does not agree 
with the set kept by such a simple ultraviolet cutoff.  In other words, 
firewalls may reside only outside the dark shaded (red) box in the 
figures, and hence may be unphysical.

We note that many of the physical states considered here do have high 
energy quanta near the horizon.  This is, however, different from the 
firewall phenomenon.  These states have many {\it physical} excitations 
near the horizon which will either fall into the black hole or fly into 
the asymptotic space within the timescale of $O(M \ln M)$; in particular, 
they cannot be obtained in the course of the standard black hole 
evaporation process.  In fact, the first two paragraphs of this section 
are sufficient to address the typicality argument for firewalls presented 
in Ref.~\cite{Marolf:2013dba}, which concerns old, near vacuum black holes.

\section{Refutation---The Entanglement Argument}
\label{sec:refute-ent}

We now address the entropy argument for firewalls.  An implicit 
assumption of the argument is that in the Hawking emission (or 
the black hole mining~\cite{Unruh:1982ic}) process, the microscopic 
information about the black hole is carried from the stretched horizon 
to the edge of the zone (or where the mining apparatus is located) by 
an excitation of a semiclassical mode:\ $B$ in Section~\ref{sec:firewall}. 
If this were indeed the case, then it would lead to a contradiction 
between unitarity and smoothness of the horizon.

The information transfer, however, does not occur in this 
manner~\cite{Nomura:2014woa,Nomura:2014voa}.  Recall that the microscopic 
information of the black hole is represented by the configuration of 
the vacuum degrees of freedom, the dark shaded (blue) box in the right 
side in the left panel of Fig.~\ref{fig:semicl-2}.  The question is how 
the black hole vacuum degrees of freedom interact with the other degrees 
of freedom:\ the modes outside the zone, $r > r_{\rm z}$, in the case 
of Hawking emission and excitation modes within the zone, $r_{\rm s} 
< r < r_{\rm z}$, in the case of mining.  The answer given in 
Refs.~\cite{Nomura:2014woa,Nomura:2014voa} is that they interact 
as if they are distributed according to the gravitational thermal 
entropy density.  This distribution is reference frame dependent, 
reflecting the fact that the vacuum degrees of freedom are not 
standard radiation, and its precise forms are not known in general. 
In a distant reference frame, however, the quasi-static nature of 
the system allows us to infer the correct distribution---the relevant 
entropy density is that obtained from the blueshifted Hawking 
temperature in Eq.~(\ref{eq:T_r}).

Since the amount of integrated entropy contained around the edge of 
the zone is of $O(1)$, outgoing field theory modes can extract the 
information {\it directly} from the vacuum degrees of freedom there, 
without involving a semiclassical mode deep in the zone.  To quantify 
this statement, we may introduce the tortoise coordinate $r^* = 
r + 2M \ln(r/2M-1)$, in terms of which the stretched horizon is at 
$r^*_{\rm s} \equiv r^*|_{r = r_{\rm s}} \simeq -4M \ln M$ and the 
edge region is $|r^*| \approx O(M)$.  We then find
\begin{equation}
  \int_{|r^*| \lesssim O(M)}\! T^3\left(r(r^*)\right) dr^* 
  \approx O(1),
\label{eq:entropy}
\end{equation}
where $T(r)$ is given by Eq.~(\ref{eq:T_r}).  This implies that the 
microscopic information about the black hole is delocalized over 
the entire zone region.%
\footnote{This is consonant with the intuition that different microstates 
 of the black hole correspond, in some sense, to black holes with 
 slightly different masses.  It is natural to expect that the information 
 about the mass is stored nonlocally, as is indeed the case classically 
 (in the form of the metric).}
Note, however, that the distribution is not uniform and is strongly 
peaked toward the stretched horizon; we obtain the full degrees of 
freedom only if we integrate the entropy density down to the stretched 
horizon
\begin{equation}
  \int_{r^*_{\rm s}}^{O(M)}\! T^3\left(r(r^*)\right) dr^* 
  \approx O({\cal A}).
\label{eq:entropy-tot}
\end{equation}
Since entropy indicates how much information one can extract from a 
system in the characteristic timescale, in this case $t \approx O(M)$, 
the amount of delocalization in Eq.~(\ref{eq:entropy}) is enough for 
outgoing field theory modes to extract an $O(1)$ amount of information 
from the vacuum degrees of freedom in each Hawking emission, which 
occurs in the timescale of $t \approx O(M)$ {\it around the edge of 
the zone}, where $t$ is the Schwarzschild time.  This is how Hawking 
emission must be viewed at the semiclassical level.  (A similar 
analysis can also be performed for the mining process.)

One might wonder what is the relation between this picture and the original 
calculation by Hawking~\cite{Hawking:1974rv}, which seems to involve 
modes deep in the zone in the semiclassical theory.  It is not uncommon 
in physics that calculation of some quantity involves ``unphysical'' 
entities in the intermediate step of the calculation.   For example, 
density fluctuations generated by cosmic inflation~\cite{Hawking:1982cz} 
are calculated by imposing the Bunch-Davies vacuum condition for all 
modes, including those that are super-Planckian at early times.  We 
do not interpret this to mean that the spacetime is indeed classical 
in sub-Planckian distances.  Likewise, a Casimir force can be 
calculated by summing up an infinite tower of modes (with an arbitrary 
large ultraviolet cutoff), and the electron anomalous magnetic moment 
can be computed by performing the momentum integral to infinity (with 
suitable counterterms).  We do not interpret these to mean that the 
theories under consideration, e.g.\ QED, are valid up to arbitrary high 
energies. In some cases, we can indeed find explicit regularizations 
which make it clear that the results do not depend on entities that 
appear in the intermediate steps of calculations.  In other cases, 
finding such explicit regularizations are difficult, but even in these 
cases, one can often be convinced (by various arguments) that the most 
naive extrapolations of the theories are giving the correct answers 
for certain ``inclusive,'' or ``low energy,'' quantities, despite the 
fact that these extrapolations cannot be taken literally.  This can 
happen because such extrapolations often capture the essential features 
of the (unknown) fundamental theories which are sufficient to guarantee 
the correct answers.  (For a discussion on an illustrative example of 
this phenomenon, see Ref.~\cite{Barbieri:2000vh}.)  We consider Hawking's 
original calculation to be of this kind---it gives the correct answers 
for the emission rate and spectrum as viewed from a distance, but we 
should not take all the intermediate steps too seriously, especially 
the part involving modes deep in the zone.

\begin{figure}[t]
\centering
  \includegraphics[clip,width=0.6\textwidth]{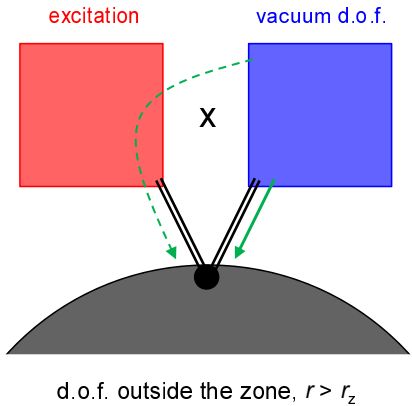}
\caption{Semiclassical degrees of freedom outside the zone, $r > 
 r_{\rm z}$, not only interact with semiclassical excitations inside 
 the zone through usual kinetic terms (the left bond) but also with 
 the vacuum degrees of freedom (the right bond).  The information 
 transfer associated with Hawking emission occurs through the latter 
 interaction (the solid arrow), rather than through semiclassical 
 excitations in the zone (the dashed arrow).}
\label{fig:outside}
\end{figure}

A schematic picture representing interactions between semiclassical 
degrees of freedom outside the zone and the black hole degrees of 
freedom (both vacuum and excitation) is given in Fig.~\ref{fig:outside}. 
The portion of the outside degrees of freedom located around $r = 
r_{\rm z}$ interact with the semiclassical degrees of freedom inside 
and near the edge of the zone through usual kinetic terms, as represented 
by the left bond in the figure.  They also interact, however, with an 
$O(1/{\cal A})$ fraction of the vacuum degrees of freedom directly, 
as indicated by the right bond.  This is where one of the assumptions 
in Ref.~\cite{Almheiri:2012rt}---the ``literal'' validity of the 
semiclassical theory outside the stretched horizon---breaks down.%
\footnote{This does not necessarily mean that generic soft quanta sent 
 to an evaporating black hole must see violation of the semiclassical 
 theory because such processes are not the same as the time reversal 
 of the Hawking emission process~\cite{Nomura:2014woa,Nomura:2014voa}. 
 To see the violation certainly, we need to send finely-tuned 
 soft quanta to an {\it anti-evaporating} black hole.  This is 
 a process in which the coarse-grained entropy decreases, since 
 usual Hawking emission is a process in which the coarse-grained 
 entropy increases~\cite{Zurek:1982zz}.}
The information transfer associated with Hawking emission occurs through 
direct interactions of the outside modes with the vacuum degrees of 
freedom (indicated by the solid arrow), rather than through semiclassical 
excitations in the zone as envisioned in the firewall argument (the 
dashed arrow).

\section{Hawking Emission:\ A Spacetime View}
\label{sec:emission}

An intuitive picture of the Hawking emission process can be obtained 
if we choose the vacuum on which excitations are defined to be the 
(hypothetical) static black hole background (the so-called Hartle-Hawking 
vacuum~\cite{Hartle:1976tp}), rather than the evolving black hole 
background as we have been doing so far.  Creation of Hawking quanta 
around the edge of the zone in this description is associated with 
that of an ingoing negative energy flux which carries {\it negative 
entropy}~\cite{Nomura:2014woa,Nomura:2014voa}.  Here, the energy and 
entropy is defined with respect to the static background.  We can 
understand this phenomenon by the following simple qubit model.

Let $\ket{\psi_k(M)}$ ($k = 1,\cdots,e^{S_0(M)}$) be the vacuum 
microstates (in the sense of the static vacuum) of the black hole 
of mass $M$.  Suppose that a black hole, in a superposition state 
of $\ket{\psi_k(M)}$'s, releases 1~qubit of information through Hawking 
emission.  This occurs in the timescale of $t \approx O(M)$, and the 
energy of the emitted quantum is $E \simeq (\ln 2)/8\pi M$, so that 
$e^{S_0(M-E)} = e^{S_0(M)}/2$.  We can model this process by saying 
that the emitted Hawking quantum is in states $\ket{r_1}$ and $\ket{r_2}$ 
if $k$ is odd and even, respectively.  Due to energy-momentum conservation, 
the process is accompanied by the creation of an ingoing negative energy 
excitation on the black hole (static) vacuum, which we denote by 
a star; namely, $\ket{\psi^*_k(M)}$ represents black hole microstates 
with the negative energy excitation.

What would this emission process look like at the microscopic level? 
Can it simply be
\begin{equation}
  \ket{\psi_k(M)} 
  \rightarrow \left\{ \begin{array}{ll}
    \ket{\psi^*_k(M)} \ket{r_1} & 
      \mbox{if $k$ is odd}, \\
    \ket{\psi^*_k(M)} \ket{r_2} &
      \mbox{if $k$ is even},
  \end{array} \right.
\label{eq:toy-1}
\end{equation}
as one might naively imagine?  If this were the case, we would find a 
problem.  Remember that $\ket{\psi^*_k(M)}$ have energy $M - E$, and 
we expect that they will relax into vacuum states of the black hole of 
mass $M - E$:
\begin{equation}
  \ket{\psi^*_k(M)} \rightarrow \ket{\psi_{k'}(M-E)}.
\label{eq:toy-2}
\end{equation}
However, since $k'$ runs only over $k' = 1,\cdots,e^{S_0(M-E)} = 
e^{S_0(M)}/2$, such a relaxation cannot occur unitarily.  Instead, 
what actually happens in the emission process is
\begin{equation}
  \ket{\psi_k(M)} 
  \rightarrow \left\{ \begin{array}{ll}
    \ket{\psi^*_{\frac{k+1}{2}}(M)} \ket{r_1} & 
      \mbox{if $k$ is odd}, \\
    \ket{\psi^*_{\frac{k}{2}}(M)} \ket{r_2} &
      \mbox{if $k$ is even},
  \end{array} \right.
\label{eq:toy-3}
\end{equation}
i.e.\ the index for the black hole microstates with the negative energy 
excitation runs only from $1$ to $e^{S_0(M)}/2$.  This allows for 
these states to relax unitarily into the black hole vacuum states 
of mass $M - E$, as in Eq.~(\ref{eq:toy-2}).  Note that the process 
in Eq.~(\ref{eq:toy-3}) is also unitary by itself if we consider the 
whole quantum state, including both the black hole and the exterior 
of the zone.

The above analysis implies that a negative energy excitation over 
the black hole static vacuum carries a negative entropy; i.e., in 
the existence of a negative energy excitation, the range over which 
the black hole microstate index runs is smaller than that without. 
Specifically, the excitation of energy $-E$ carries entropy $-8\pi M E$. 
This picture is rather comfortable, since entropy is usually associated 
with energy, $S \sim E$, and we are saying that this is also the 
case even if these quantities are measured with respect to the static 
black hole background.  We find that the information transfer from 
an evaporating black hole occurs through an ingoing negative entropy 
flux, at least from this viewpoint.

A comment is in order.  Since the creation of a Hawking quantum, and 
hence of a negative energy excitation, occurs in the timescale of $O(M)$, 
and the relaxation time of a negative excitation is expected to be 
of $O(M \ln M)$, the amount of negative energy excitations we have 
on the static black hole background is of order $\ln M$ at any time. 
Here, the relaxation timescale can be estimated from the time it 
takes for the excitation to propagate from the edge of the zone to 
the stretched horizon and the time it takes for the information to 
be scrambled~\cite{Hayden:2007cs}, both of which give $O(M \ln M)$. 
We may therefore view that an evaporating black hole has steady negative 
energy and entropy {\it fluxes} and redefine the black hole vacuum 
to include them.  The resulting vacuum then has entropy $S(M)$, given 
by $S(M)-S_0(M) \approx -\ln M$.  This redefined vacuum corresponds, 
very roughly, to the Unruh vacuum~\cite{Unruh:1976db} in the semiclassical 
theory, and the corresponding geometry is that of an evaporating black 
hole, which is well described by the advanced/ingoing Vaidya metric 
near the horizon~\cite{Bardeen:1981zz}.  In this picture, the change 
in the local gravitational field supplies the energy of outgoing Hawking 
quanta created at $r \approx r_{\rm z}$.  The dark shaded (blue) boxes 
in the right side in (b), (c), (d) of Fig.~\ref{fig:semicl-1} and in 
the left panel of Fig.~\ref{fig:semicl-2} represent the microscopic 
degrees of freedom associated with this redefined vacuum.

\begin{figure}[t]
\centering
  \subfigure{\includegraphics[clip,width=.49\textwidth]{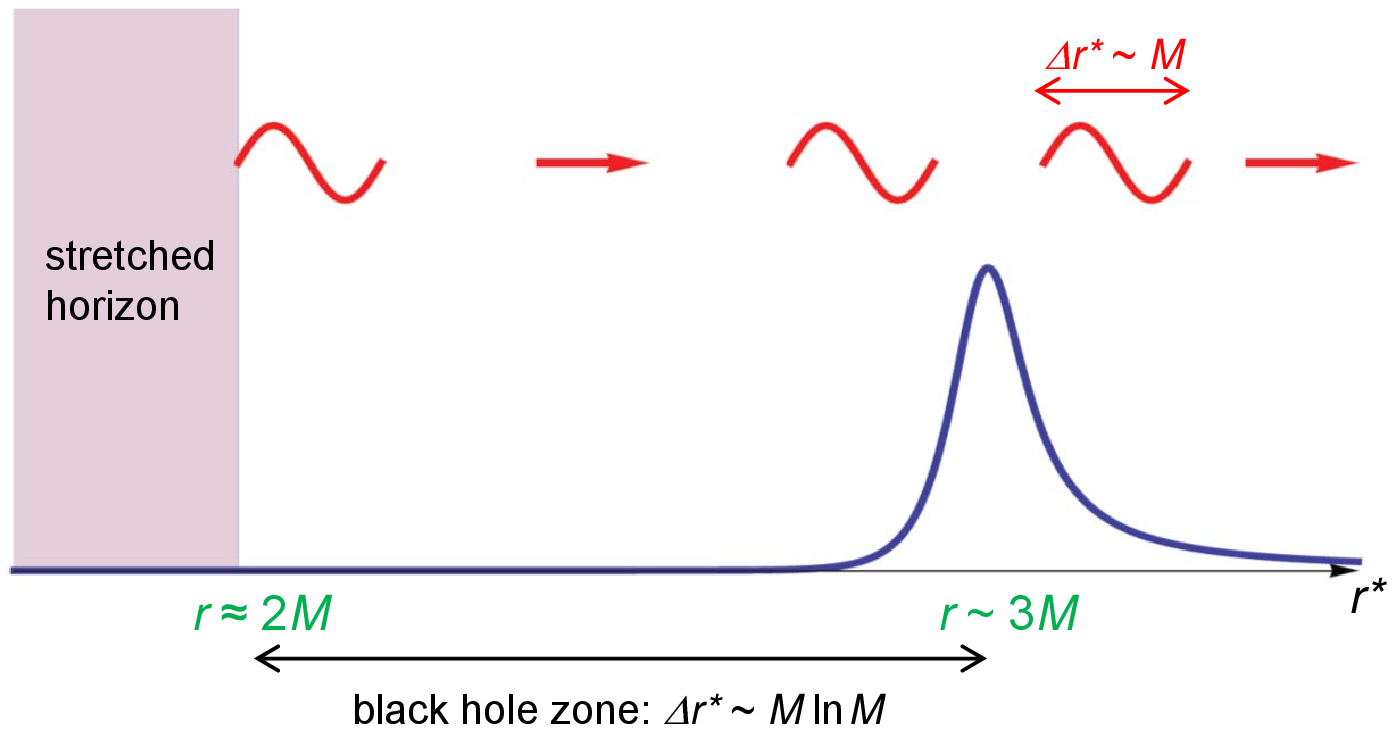}}
  \subfigure{\includegraphics[clip,width=.49\textwidth]{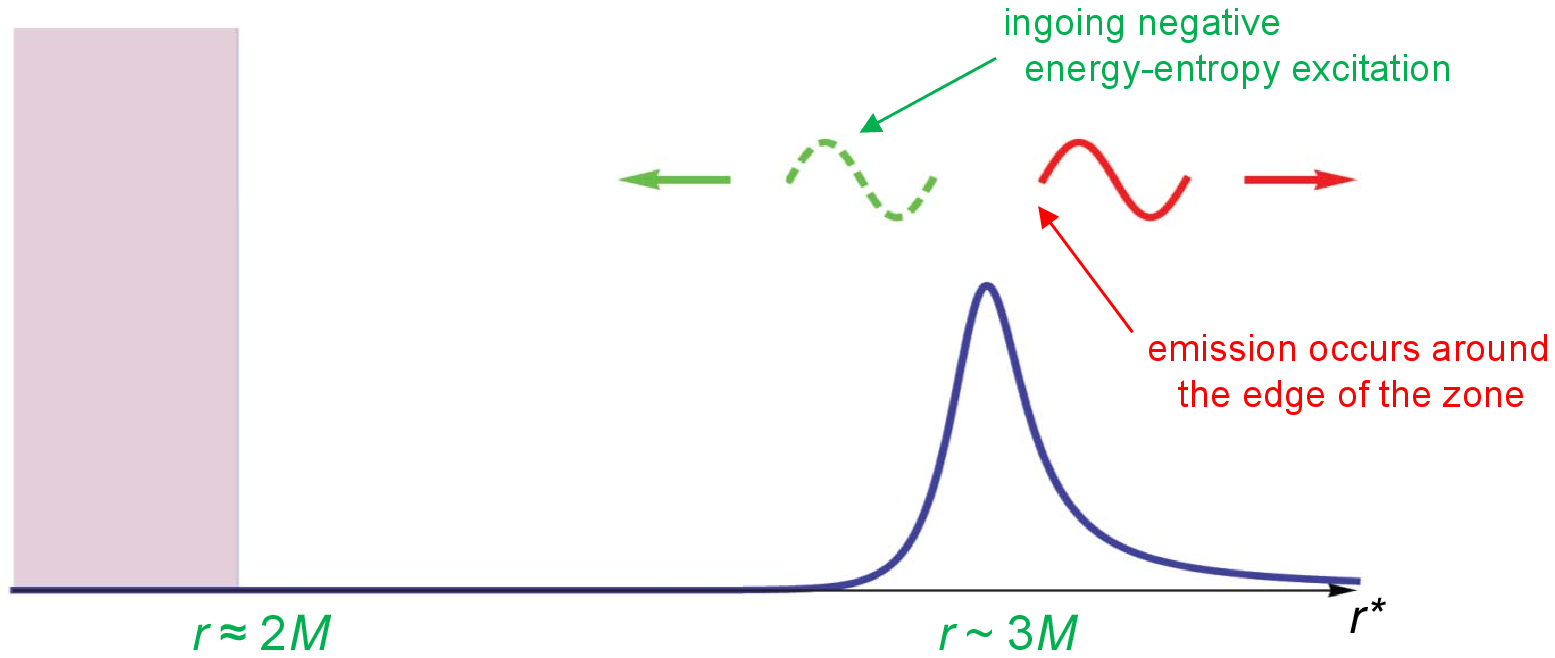}}
\caption{The information transfer from an evaporating black hole does 
 not occur through outgoing positive energy-entropy excitations (left 
 panel) but through ingoing negative energy-entropy excitations in 
 the zone (right panel).  This is possible because the microscopic 
 information about the black hole is carried by the ``spacetime itself'' 
 (the vacuum degrees of freedom), which at the semiclassical level 
 must be viewed as delocalized over the zone according to the thermal 
 entropy density associated with the blueshifted Hawking temperature.}
\label{fig:info-trans}
\end{figure}

The picture of Hawking emission resulting from the above 
analysis~\cite{Nomura:2014woa,Nomura:2014voa} is different from 
what was imagined in Refs.~\cite{Almheiri:2012rt,Marolf:2013dba,%
Almheiri:2013hfa,Polchinski:2015cea}, which implicitly assumed that 
some information transportation mechanism is in operation from the 
stretched horizon to the edge of the zone {\it on the semiclassical 
background}; see the left panel of Fig.~\ref{fig:info-trans}.
Our picture says that the information transfer from an evaporating black 
hole cannot be understood in this manner---it is the {\it spacetime 
itself} that carries the microscopic information about the black hole, 
and this information must be viewed as delocalized throughout the zone 
in the semiclassical picture.  With respect to the static background, 
the transfer occurs through an ingoing flux of negative energy-entropy 
excitations created around the edge of the zone, as depicted in the right 
panel of Fig.~\ref{fig:info-trans} (although these excitations can be 
incorporated as a part of the evolving black hole vacuum).  The absence 
of the problem found in the entanglement argument is now obvious:\ 
there is no outgoing mode that is entangled with both early radiation 
and the mirror mode.  While late Hawking quanta are certainly entangled 
with early ones for an old black hole, these quanta exist only outside 
the zone, where the near horizon approximation is not applicable (and 
hence there is no such thing as the mirror modes).

Comparing this picture~\cite{Nomura:2014woa,Nomura:2014voa} with the 
old, heuristic picture of Hawking's pair creation~\cite{Hawking:1974rv,%
Hawking:1976ra}, we find two key features which we reiterate here:
\begin{itemize}
\item
From the semiclassical viewpoint, the location in which pairs of a 
positive energy Hawking quantum and a negative energy excitation are 
created is {\it not} at the (stretched) horizon but around the edge 
of the zone, which is {\it macroscopically} away from the horizon.%
\footnote{Similar points have also been discussed more recently in 
 Refs.~\cite{Israel:2015ava,Giddings:2015uzr}.}
Microscopic information about the black hole is transferred there to 
field theory quanta, as in Eq.~(\ref{eq:toy-3}), which is possible 
because the information is carried by the spacetime itself and so 
is delocalized over the entire zone region.  Note that it is not 
unnatural for such special dynamics to occur in this particular region, 
since it is where the near horizon, Rindler-like space is ``patched'' 
to the asymptotic, Minkowski-like space.
\item
The creation of a positive energy Hawking quantum and a negative energy 
excitation takes a form very different from the standard ``pair creation'' 
of particles.  In the standard pair creation picture, the final states 
associated with the positive and negative energy excitations are assumed 
to be maximally entangled with each other, which is {\it not} the 
case here as one can see by writing explicitly the expression in 
Eq.~(\ref{eq:toy-3}) for the first few $k$'s.  For example, the black 
hole states after the emission are the same for $k = 1$ and $2$, despite 
the fact that the states for the emitted quanta are different.  In fact, 
it is this lack of entanglement that allows for the emission process 
to transfer the information from the black hole to the radiation.
\end{itemize}
The calculation by Hawking ``bypasses'' these points while still giving 
the correct answers for the rate and spectrum of the emitted quanta 
as viewed from a distance.  This must be because it captures an essential 
feature(s) of the fundamental theory, which is ultimately responsible 
for this energy-information transfer process between spacetime and 
particles.

What is the essential feature Hawking's calculation is capturing?  We 
suspect that it may exactly be the smoothness of the horizon, i.e.\ 
the ability of erecting a reference frame in which physics looks 
approximately Minkowskian locally there.  Hawking's (or other related) 
calculation provides an effective way of incorporating this information 
into the derivation of the rate and spectrum of the emitted particles. 
As we have argued, while we may trust these quantities as viewed from 
a distance (or from ``high energy'' excitations such as a mining detector 
in the zone) since the black hole physics is already ``integrated out,'' 
it does not mean that all the intermediate steps of the calculation 
can necessarily be trusted.  To diagnose if we can in analogous cases, 
we usually analyze if the naive interpretation of the theory leads 
to pathological conclusions.  In some cases these pathologies are readily 
evident, but in general not all sicknesses of effective theories are 
straightforward to see (cf.~Ref.\cite{Adams:2006sv}).  The firewall 
arguments may be viewed as such a pathology, indicating the limitation 
of the semiclassical theory interpreted naively.

\section*{Acknowledgments}

This work was supported in part by the Director, Office of Science, 
Office of High Energy and Nuclear Physics, of the U.S.\ Department of 
Energy under Contract DE-AC02-05CH11231, the National Science Foundation 
under grants PHY-1521446, and MEXT KAKENHI Grant Number 15H05895.

\end{document}